\documentclass[a4paper,12pt,oneside,english]{report}
\usepackage{natbib} 
\usepackage{mathptmx}
\usepackage{amsmath}
\usepackage{amsfonts}
\usepackage{amssymb}
\usepackage{rotating}
\usepackage{subfigure}

\usepackage{graphicx}
\usepackage{babel}
\usepackage{epsfig}

\def\nat{{\it Nature,}}
\def\jgr{{\it Journal of Geophysical Research (Space Physics)}}
\def\grl{{\it Geophys. Res. Lett.,}}
\def\mnras{{\it Mon. Not. R. Astron. Soc.}}
\def\apj{{\it Astrophys. Jou.}}

\def\aap{{\it Astron. \& Astrophys.}}

\def\solphys{{\it Sol. Phys.}}

\makeatother
\begin{document}
\noindent Accepted for publication in {\bf{\textit{Geophys. Res. Lett. - 2011}}}\\
\line(1,0){490}
\vspace{0.5cm}\\\\
{\LARGE{\bf{\noindent The Prelude to the Deep Minimum between Solar Cycles 23 and 24: Interplanetary 
Scintillation Signatures in the Inner Heliosphere}}}

\begin{center}
P. Janardhan${^1}$, Susanta Kumar Bisoi${^1}$, S. Ananthakrishnan${^2}$, \\ M. Tokumaru${^3}$ and K. Fujiki${^3}$
\end{center}
$\;\;$ \\
\noindent ${^1}$ {{Physical Research Laboratory, Ahmedabad 380 009, India. \\email: jerry@prl.res.in email: susanta@prl.res.in}} \\\\
\noindent ${^2}$ {{Electronic Science Department, Pune University, Pune 411 007, India. \\email: subra.anan@gmail.com}} \\\\
\noindent ${^3}$ {{Solar-Terrestrial Environment Laboratory, Nagoya University, 
Honohara 3-13, Toyokawa, Aichi, 442-8507, Japan. \\email: tokumaru@stelab.nagoya-u.ac.jp email: fujiki@stelab.nagoya-u.ac.jp}} \\

\section*{Abstract}
{\it{Extensive interplanetary scintillation (IPS) observations at 327 MHz 
obtained between 1983 and 2009 clearly show a steady and significant 
drop in the turbulence levels in the entire inner heliosphere starting 
from around $\sim$1995. We believe that this large-scale IPS signature, 
in the inner heliosphere, coupled with the fact that solar polar fields 
have also been declining since $\sim$1995, provide a consistent result 
showing that the buildup to the deepest minimum in 100 years actually 
began more than a decade earlier.}}

\section*{Introduction}\label{S-Intro} 
The sunspot minimum at the end of Cycle 23, has been one of the deepest 
we have experienced in the past 100 years with the first spots of the new cycle 
24 appearing only in March 2010 instead of December 2008 as was expected.  Also, 
the number of spotless days experienced in 2008 and 2009 was over 70\%.  
Apart from this, Cycle 23 has shown a slower than  
average field reversal, a slower rise to maximum than other odd numbered cycles, 
and a second maximum during the declining phase that is unusual for odd-numbered 
cycles.  Though these deviations from ``normal'' behaviour could be significant 
in understanding the evolution of magnetic fields on the Sun, they do not 
yield any direct insights into the onset of the deep minimum experienced 
at the end of cycle 23.  This is because predictions of the strength of solar 
cycles and the nature of their minima are strongly dictated by both the strength 
of the ongoing cycle [\cite{DTG06, CCJ07}] and changes in the flow rates of the 
meridional circulation [\cite{NMM11}]. 

Ulysses, the only spacecraft to have explored the mid- and high-latitude heliosphere, 
in its three solar orbits, provided the earliest indications of the global changes 
taking place in the solar wind.  A significant result from the Ulysses mission  
came from observations of $\left|B{_{r}}\right|$, the radial component 
of the interplanetary magnetic field (IMF), as a function of the heliographic
distance r which showed that the product $\left|B{_{r}}\right|r{^{2}}$ is 
independent of the heliographic latitude [\cite{SmM03,LOR09}].  This
result has far reaching consequences in that it essentially enables one to use 
{\sl{in~situ}}, single-point observations to quantify the open solar
flux entering the heliosphere.  The second Ulysses orbit, covering the rising to
maximum phase of cycle 23, found the solar wind dynamic pressure (momentum flux) to 
be significantly lower in the post maximum phase of cycle 23 than during its earlier 
orbit, spanning the declining phase of cycle 22 [\cite{RiW01, McE03}]. Finally, the 
third Ulysses orbit (2004 -- 2008) found a global reduction of open magnetic flux 
and showed an $\sim$ 20\% reduction in both solar wind mass flux and dynamic 
pressure in cycle 23 as compared to the earlier two cycles [\cite{McE08}]. 
In addition, a study using solar wind measurements between 
1995--2009 [\cite{JRL11}] has shown that the  solar minimum in 2008--2009 has 
experienced the slowest solar wind with the weakest solar wind dynamic 
pressure and magnetic field as compared to the earlier 3 cycles.
%
\protect\begin{figure}[ht]
\vspace{7.5cm}
\includegraphics{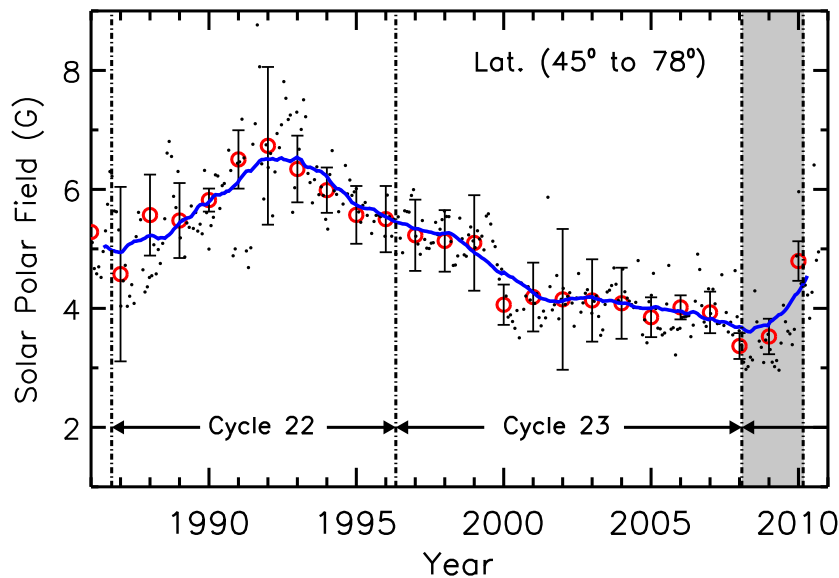}
\caption{Shows the absolute value of the solar polar field, in the latitude 
range 45 --78${^{\circ}}$, as a function of time in years for solar cycles 22 
and 23.}
\label{mag-field}
\end{figure}	  
%
Figure 1 shows the solar magnetic field in the latitude range 
45${^{\circ}}$ --78${^{\circ}}$ computed from ground based magnetograms.  
The filled dots in Figure 1 represent the actual measurements 
and the open circles are the yearly means with 1$\sigma$ error bars.  The solid 
line is a smoothed curve. The vertically oriented dashed parallel lines 
demarcate cycles 22, and 23, respectively and the shaded gray area indicates 
the time between the expected minimum in cycle 23 and the time when the first 
spots in cycle 24 actually began to appear. 

The method adopted in computing solar magnetic fields 
shown in Figure 1 and details of the database used has been 
described in [\cite{JBG10}]. It is clear from Figure 1 that 
there has been a continuous decline in the magnetic field starting from 
around 1995.  Since the IMF is basically the result of photospheric magnetic 
fields being continuously swept out into the heliosphere one would expect 
to see this reflected in the solar wind and interplanetary medium.  Due to 
the fact that the solar wind undergoes an enormous change in densities ranging 
from $\sim$10${^{9}}$ cm${^{-3}}$ at the base of the corona [\cite{MaK03}] 
to $\sim$10 cm${^{-3}}$ at 1 AU, different techniques are needed to study 
different regions of the solar wind. However, IPS is the only technique that 
can probe the entire inner-heliosphere  using ground based radio telescopes 
operating at meter wavelengths.   
\section*{The IPS Methodology}

IPS is a scattering phenomenon in which one observes distant extragalactic 
radio sources to detect random temporal variations of their signal 
intensity (scintillation) which are caused by the scattering suffered when 
plane electromagnetic radiation from the radio source passes through the 
turbulent and refracting solar wind [\cite{HSW64, ACK80, AsK98, Man10, TKF10}].  
Though IPS measures only small-scale ($\sim$150 km sized) fluctuations in density 
and not the bulk density itself, \cite{HTG85} showed that there was no 
evidence for enhanced or decreased IPS that was not associated with corresponding 
variations in density.  They went on to derive a relation between a normalized 
scintillation index denoted `g' and the density given by 
g=(Ncm${ ^{-3}}$/9)${^{0.52\pm 0.05}}$. 
Thus, whenever interplanetary disturbances, containing either enhanced or 
depleted rms electron density fluctuations ($\Delta$n${_{e}}$) as compared to the 
background solar wind, cross the line-of-sight (LOS) to the 
observed source they exhibit themselves as changes in the levels of scintillation 
(m), {\it{i.e.}} higher or lower than expected m, where 
m = ${{\Delta S }/{\left\langle S \right\rangle}}$ is the ratio of the scintillating 
flux $\Delta$S to the mean source flux $\left\langle S \right\rangle$.  In other words, 
whenever turbulence levels change in the solar wind, they will be reflected in IPS 
measurements as changes in m.  The main advantage of the IPS technique is that, it 
can probe a very large region of the inner heliosphere, ranging 
from about 0.2 AU to 0.8 AU and it is extremely sensitive to small changes in 
$\Delta$n${_{e}}$.  In fact IPS is so sensitive to changes in $\Delta$n${_{e}}$ 
that it has been used to probe $\Delta$n${_{e}}$ fluctuations in tenuous cometary 
ion tails well downstream of the nucleus [\cite{JaA92}] and to study solar 
wind disappearance events wherein average densities at 1 AU drop to values below 
0.1 cm${^{-3}}$ [\cite{JaF05}].  In a typical IPS observation, the angle
between the Sun the Earth and the observed radio source is known as the solar 
elongation ($\epsilon$).  Please refer to Figure 1 of [\cite{BaJ03}] for details 
of the IPS observing geometry.  Since the solar wind density beyond $\sim$0.2AU 
is inversely proportional to r${^{2}}$ [\cite{BiV94}] where, 
r~=~sin($\epsilon$), measurements of m in the direction of a given radio 
source will increase with decreasing $\epsilon$ or distance 'r' from the 
Sun, until a certain $\epsilon$.  After this point m falls off sharply with 
further reduction in $\epsilon$.  The $\epsilon$ at which m turns over is 
dependent on frequency and at 327 MHz it lies between 10${^\circ}$ and 14${^\circ}$. 
The region at $\epsilon$ larger than the turnover defines 
the region of weak scattering where the approximation of scattering by 
a thin screen is valid [\cite{Sal67}]. For an ideal point source, m  
will be unity at an $\epsilon$ of $\sim$12${^{\circ}}$ 
and drop to values below unity with increasing distance r.  For sources, 
with finite compact component sizes ($>$ 0 milli arcsec (mas)), 
m will be $<$ unity at the turnover.

For the steady state solar wind, values of m can be computed by 
obtaining theoretical temporal power spectra using a solar wind model 
assuming weak scattering and a power law distribution of density irregularities 
in the IP medium for any given source size and distance r of the LOS from 
the Sun [\cite{Mar75}].  
%
\protect\begin{figure}[ht]
\vspace{6.5cm}
\includegraphics{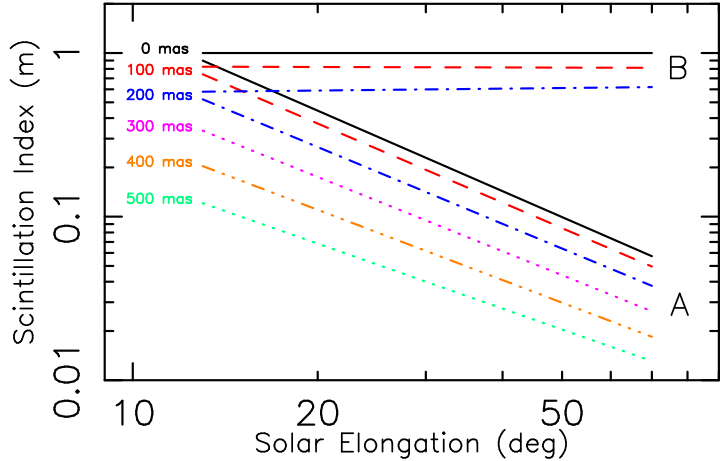}
\caption{Curves (labeled 'A') of theoretically expected m as a 
function of $\epsilon$ for various source sizes. Curves (labeled 'B') 
corresponding to source sizes of 0 mas, 100 mas, and 
200 mas, have been normalized by the point 
source m at the corresponding $\epsilon$.}
\label{marians}
\end{figure}	  
%
The set of six curves (labeled 'A') in Figure 2 
show the theoretically expected m as a function of $\epsilon$ for various 
source sizes in mas, assuming weak scattering at 327 MHz. To 
remove the $\epsilon$ or distance dependence of m, each observation of m has to 
be normalized by that of a point source at the corresponding $\epsilon$.  It has 
been shown that the source 1148-001 is $\le$10 mas in angular extent at 327 MHz.  
Thus, 1148-001 can be treated as a nearly ideal point source, with most of its flux 
contained in a scintillating compact component [\cite{VeA85}]. Normalizing all 
observations of m in this manner will yield an $\epsilon$ or distance independent 
value of m as shown for three source sizes (labeled 'B') in Figure 2.  
It must be noted that the curves in Figure 2 imply that the radial 
fall off rate is slightly different for different source sizes.  Thus, the m 
for the larger source sizes will be over-corrected at large distances.  However, 
most of the observed IPS sources are strong scintillators with sizes $\le$250 mas 
and will not suffer from this normalization.  

\section*{The Observations and Data Reduction}

The three-station IPS facility, operated by the Solar-Terrestrial Environment 
Laboratory (STEL), Japan, and used for the 
observations described in this paper, can carry out IPS observations 
at 327 MHz on about 200 compact extragalactic, radio sources on a regular 
basis, to derive solar wind velocities and m.  In 1994, a fourth antenna was
added to the system to form a four-station network that could provide 
more robust estimates of the solar wind speed owing to the redundancy 
in the baseline geometry obtained by having an additional station. We 
have taken data from 1983 to 2009, which had a one year data gap in 
1994 owing to the development of the fourth IPS station, and have chosen
sources which had at least 400 individual observations over this 
period of 27 years.  We also ensured that there were no significant 
data gaps in any given year, apart from the one year gap in 1994. 
Of the total of 215 sources observed, we were finally left with 26 sources
covering the entire 24 hour range of source right ascensions and 
a wide range of source declinations.  The source 1148-001 was included 
as the 27th source even though it had some data gaps and therefore did not
fully comply with our selection criteria.  The data set was first normalized 
by the highest value so that the range of m was between 0 and unity.  The 
distance dependence of m was then removed by normalizing every individual 
observa-
\protect\begin{figure}
\vspace{20.0cm}
\includegraphics{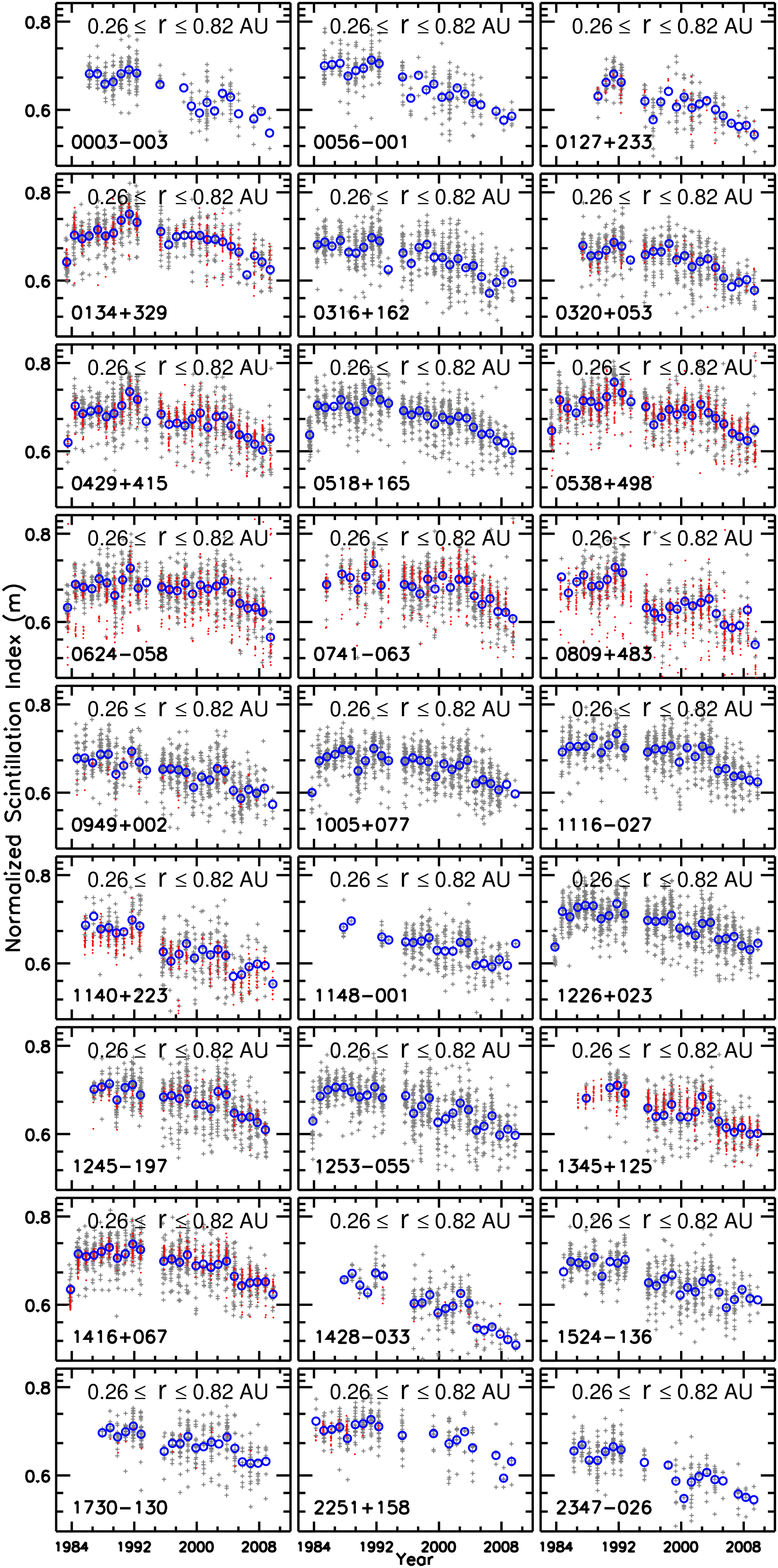}
\caption{Plots m as a function of time in years 
for 27 sources after m has been made independent 
of distance from the Sun by normalizing each 
observation by the value of m 
for the source 1148-001.  The grey 
crosses are observations at source helio latitudes 
$\leq$45${^{\circ}}$ while the fine red dots are 
observations at source heliolatitudes $>$45${^{\circ}}$.  
The open circles are yearly averages by excluding the 
high latitude observations.  The IAU name of each 
source is indicated at the bottom left in each panel.}
\label{sindex}
\end{figure}	    
%
\noindent tion for each source by the value of m for the source 1148-001 at 
the corresponding $\epsilon$.  Figure 3 shows plots of m as a function 
of time in years for the 27 chosen sources lying in the distance range 0.26 to 0.82 AU.  
The grey crosses are measurements when the source heliolatitude is $\leq$ 45${^{\circ}}$ 
while the fine red dots are measurements when the source heliolatitude is 
$>$45${^{\circ}}$.  The large, blue, open circles are yearly averages taken 
without the high latitude observations.The reason for dropping the high 
latitude observations from the yearly averages is because IPS observations 
have shown [\cite{ToK00}] that the solar wind structure changes with
the solar cycle, being more or less symmetric at solar maximum and considerably 
asymmetric during solar minimum.  This asymmetry could change the way m falls 
off with radial distance.  Since our IPS measurements mix measurements distant 
from the Sun near the solar equator with measurements at small $\epsilon$ at 
high solar latitude, we drop the high latitude observations from the yearly 
averages in order not to affect the yearly means. In addition, long term and 
gradual changes in the antenna sensitivity caused by degradation in 
the efficiency of the system, could cause some changes in the IPS 
measurements over time.  Such changes are very difficult to quantify but every 
effort has been made to maintain the system stability and we believe that 
these changes cannot account for or explain the systematic drop in m starting 
from around 1995.  It can be seen from Figure 3 that all the sources 
show a steady decline in m starting from around $\sim$1995 implying a reduction 
in solar wind turbulence levels.  The drop in m ranges between 10\% and 25\% when 
the annual means are taken considering all observations and it ranges between 10\% 
and 22\% when the annual means are taken without including the high latitude 
observations.  The average drop is therefore around 16\%.  One must note 
however that only 16 of the 27 sources go to latitudes above 45${^{\circ}}$ 
while the remaining 11 sources have all observations at heliolatitudes 
$\leq$45${^{\circ}}$.  
%
\protect\begin{figure}[ht]
\vspace{8.0cm}
\includegraphics{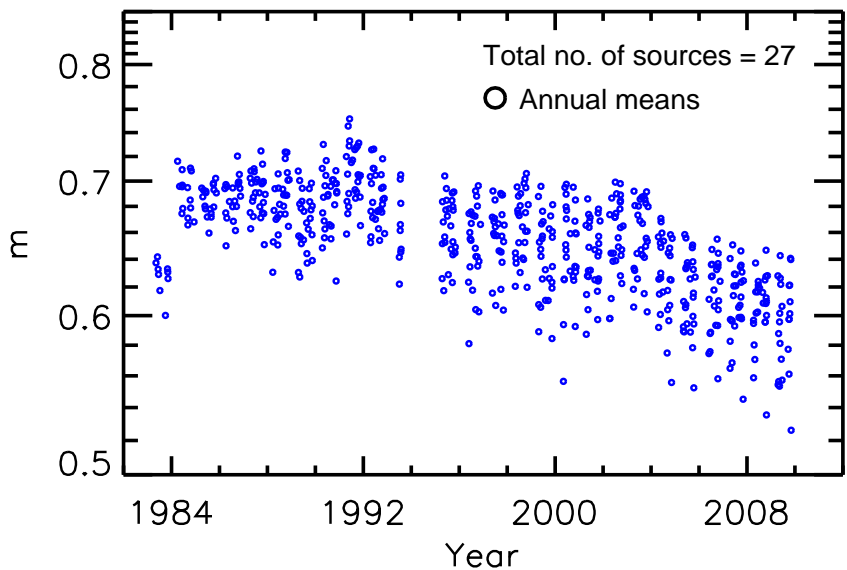}
\caption{Shows the annual means of m as a function 
of time for each of the the 27 sources shown in 
Fig. 3.}
\label{yearly-means}
\end{figure}	  
%
In order to achieve greater clarity and to avoid any confusion caused by 
showing a large number of panels in Figure 3 the yearly means 
of each source (blue open circles) are shown again, in Figure 4, 
as a composite plot of m as a function of time in years.  The steady drop in 
m from $\sim$1995 can be unambiguously seen in Figure 4. 

\section*{Discussion and Conclusions}

We have seen that the turbulence levels, as indicated by reduced scintillation 
levels in the IP medium, have dropped steadily since $\sim$ 1995.  In addition, 
$\left|B{_{r}}\right|$ has been lower during the recent minima than in the past 
three minima [\cite{SBa08}] and solar polar fields have also shown a steep decline 
since $\sim$1995.  Since solar polar fields supply most of the heliospheric magnetic 
flux during solar minimum conditions [\cite{SCK05}], weaker polar fields imply that the 
IMF will also be significantly lower.  A causal relationship between stronger 
magnetic fields and turbulence in this region also implies a decrease in turbulence 
levels over the solar poles since $\sim$1995.  It has been shown that the solar 
wind turbulence is related to both the rms electron density fluctuations 
($\Delta$n${_{e}}$) and large scale magnetic field fluctuations in fast solar wind
streams [\cite{ACK80}].  Similarly, it is possible that a global reduction in the 
IMF is being reflected as a large scale global reduction in m (or microturbulence)
as shown by our observations.  As opposed to our method, IPS observations of the
change in the turnover distance of m, which defines the minimum distance (from 
the sun) at which the weak scattering regime starts, have shown a steady decrease
in the scattering diameter of the corona since 2003 [\cite{Man10a}]. 

A great deal of work has been done in modeling the solar dynamo and 
predicting the strength of future solar cycles.  It has been argued 
[\cite{DTG06}] that since the meridional circulation period is between 
17--21 years in length, the strength of the polar fields during the 
ongoing cycle minimum and the preceding two solar minima will influence 
the strength of the next cycle.  Another view point is that of 
[\cite{CCJ07}] who claim that only the value of the polar field strength 
in the ongoing cycle minimum is important in predicting the next cycle.  
Very recently, [\cite{NMM11}] used a kinematic dynamo simulation model that 
adjusted the flow speeds in each half of the cycle to reproduce 
the extended minimum in cycle 23.  Their model showed that the changes in the
meridional flow speeds that led to the extended minimum began as 
early as the mid to late 1990's and also predicted that very deep 
minima are generally associated with weak polar fields. On the other hand, 
using observations, from cycles 21, 22 and 23, of the drift of Fe XIV 
emission features over time at high latitudes, referred to as ``rush to the 
poles'' [\cite{Alt11}] was able to show that the solar maximum occurs
approximately 1.5 years prior to the rush-to-the-poles reaching the solar poles. 

In conclusion, our observations have shown that turbulence levels 
in the inner heliosphere have shown a steady decline since the mid 90's.  
The fact that the change in the meridional flow that regulates 
the solar dynamo began in the mid 1990's coupled with the fact that solar polar 
fields have been declining since $\sim$1995 and our IPS observations all provide 
a consistent result showing that the buildup to the deepest minimum in 100 
years actually began more than a decade earlier.

\section*{acknowledgments}
{\it{IPS observations were carried out under the solar wind program of STEL, Japan. 
We thank the two anonymous referees whose comments have improved the paper 
significantly. One of the authors (JP) thanks Bill Coles, Mike Bird and 
Murray Dryer for their critical comments.}}

\end{document}